\documentclass[reprint,amssymb, amsmath, aip,cha,twocolumn]{revtex4-1}
\usepackage{graphicx}
\usepackage[caption = false]{subfig}
\usepackage{color}
\usepackage{epsfig}
\usepackage{ifpdf}
\usepackage{url}%
\usepackage{amsmath}
\usepackage{array}
\usepackage{float}
\usepackage{bm}
\usepackage[colorlinks=true,linkcolor=blue]{hyperref}%
\usepackage{cleveref}
\usepackage{ulem}
\expandafter\ifx\csname package@font\endcsname\relax\else
\expandafter\expandafter
\expandafter\usepackage
\expandafter\expandafter
\expandafter{\csname package@font\endcsname}%
\fi
\begin{document}
	\title{Effect of polarization on spectroscopic characterization of laser produced aluminium plasma }%
	\author{B R Geethika}%
	\email{geethika.br@ipr.res.in}
	\affiliation{Institute For Plasma Research, Bhat, Gandhinagar, Gujarat, 382428, India}%
	\affiliation{Homi Bhabha National Institute, Training School Complex, Anushaktinagar, Mumbai, 400094, India}%
	
	\author{Jinto Thomas}
	\email{jinto@ipr.res.in}
	\affiliation{Institute For Plasma Research, Bhat, Gandhinagar, Gujarat, 382428, India}%
	\affiliation{Homi Bhabha National Institute, Training School Complex, Anushaktinagar, Mumbai, 400094, India}%
	\author{Renjith Kumar R}
	\affiliation{Institute For Plasma Research, Bhat, Gandhinagar, Gujarat, 382428, India}%
	\affiliation{Homi Bhabha National Institute, Training School Complex, Anushaktinagar, Mumbai, 400094, India}%
	\author{Janvi Dave}
	\affiliation{Institute For Plasma Research, Bhat, Gandhinagar, Gujarat, 382428, India}%
	\author{Hem Chandra Joshi}
	\affiliation{Institute For Plasma Research, Bhat, Gandhinagar, Gujarat, 382428, India}%

	\date{\today}
	\begin{abstract}
		Laser-induced breakdown spectroscopy (LIBS) is a well-established technique widely used in fundamental research and diverse practical fields. Polarization-resolved LIBS, a variant of this technique, aims to improve the sensitivity, which is a critical aspect in numerous scientific domains. 
		In our recent work we demonstrated that the degree of polarization (DOP) in the emission depends on the spatial location and time in a nano second laser generated aluminium plasma\cite{geethika}. Present study investigates the effect of polarized emission on the estimation of plasma parameters. The plasma parameters are estimated using the conventional spectroscopic methods such as Boltzmann plot and line intensity ratio for the estimation of electron temperature and Stark broadening for estimating the electron density. The estimated plasma temperature 
		using Boltzmann plot method shows large errors in electron temperature for the locations where DOP is higher. 
		However, the electron density estimated using the Stark width does not show such variation. 
		The observed ambiguity in temperature estimation using the Boltzmann plot method appears to be a consequence of deviation from expected Maxwell Boltzmann distribution of population of the involved energy levels.
		 These findings highlight the need of assessing the DOP of the plasma before selecting the polarization for PRLIBS or temperature estimation using Boltzmann plots in elemental analysis.
		 
	\end{abstract}
	\maketitle
	\section{Introduction}\label{sec:intro}

	Laser-Produced Plasma (LPP) is a dynamic system comprises of intricate processes.The interaction between high-intensity laser and a material induces the formation of plasma, characterized by emissions specific to the constituent elements of the material. Laser-Induced Breakdown Spectroscopy (LIBS)\cite{libs_basic,libs_general} exploits these characteristic line emissions to discern and quantify the elemental composition of the material. LIBS is a valuable tool for elemental analysis encompassing various applications\cite{IDRIS,pitchstone_iceland,archeology,nails_libs}. Several variants in LIBS technique had been materialized in order to enhance its sensitivity\cite{rev_jinto,sensitivity}. Polarization-Resolved Laser-Induced Breakdown Spectroscopy (PRLIBS), a modified version of LIBS technique used to enhance detection limit by improving signal to noise ratio\cite{liu_2008,Zhao_2014,TakashiFujimoto_1999}, choosing single polarization of the line emission over un-polarized background emission.
	\par
	Polarized emission from LPP is believed to be the result of distinct elementary processes within the plasma, as outlined by various research groups\cite{AghababaeiNejad2017,SHARMA20073113}. Our earlier study\cite{geethika} demonstrated that the degree of polarization (DOP) depends on both time elapsed after ablation and spatial positions within the plasma plume. More importantly, it showed that the DOP flip its sign as the distance from sample increases.  The cause of this polarized emission is qualitatively stated as the effect of self generated magnetic field\cite{SHARMA20073113,geethika}.
	This observation holds considerable importance in the context of PRLIBS as the selection of particular polarization needs special attention to enhance the signal to noise ratio. Selection of a polarization without the knowledge of DOP at a particular position and time, can be detrimental, in improving the signal to noise ratio as anticipated in PRLIBS.
	\par
	Electron density and temperature of the plasma are important parameters to quantify the elemental composition of a material using LIBS analysis\cite{cf_libs,cf_libs2}. Optical Emission Spectroscopy (OES) has been used to estimate electron number density\cite{densharilal,rev_jinto} and temperature\cite{harilaltemp,ARAGON2008893,khudtson_1987} in LPP. For plasmas with high electron density, as in case of LPP, Stark broadening is widely exploited for accurate electron density determination. However, despite of the potency of OES in LPP characterization, it relies on certain specific assumptions\cite{ARAGON2008893}. Most important assumption is that the plasma should hold condition of local thermodynamic equilibrium (LTE)\cite{ARAGON2008893,CRISTOFORETTI201086}, where collisional processes dominate over radiative ones\cite{grieim} and all the species present within the plasma are in thermal equilibrium. In such a scenario the electron temperature is estimated by Boltzmann plot method\cite{AGUILERA_bplot,ARAGON2008893,harilaltemp}. Another method which is adopted to calculate electron temperature is the line intensity ratio from successive ionic states\cite{grieim,densharilal} which is known to be less error prone compared to the Boltzmann plot method.
	\par
	As mentioned, the applicability of Boltzmann plot method strongly depends on the validity of LTE. Previous studies were aimed to find potential causes of inaccuracy of the method when the excitation mechanism of populating the higher energy levels is non-thermal\cite{ZHANG} as well as precautions that had to be implemented while employing this approach\cite{BOUSQUET}.
	Despite these challenges, the method perseveres as a prevalent means of electron temperature measurement, particularly within the domain of LIBS for elemental analysis. 
	\par
	Our earlier article\cite{geethika}, clearly demonstrated that the laser produced plasma of aluminium at higher background pressures exhibits large anisotropy with significant variation along the plasma plume propagation direction. 
	In this work, we study the effect degree of polarization (DOP) on the estimation of plasma parameters, such as electron density and temperature. Our findings highlight the importance of estimation of DOP for line emissions before using the Boltzmann plot method to estimate electron temperature. This is crucial as the Boltzmann plot method is extensively used in LIBS and PRLIBS studies for composition analysis of materials.

	\section{Experimental Set-up}\label{sec:setup}
	
	\begin{figure}[h!]
		\includegraphics[scale = 0.3,trim = {0.8cm 1.5cm 0 0}, clip]{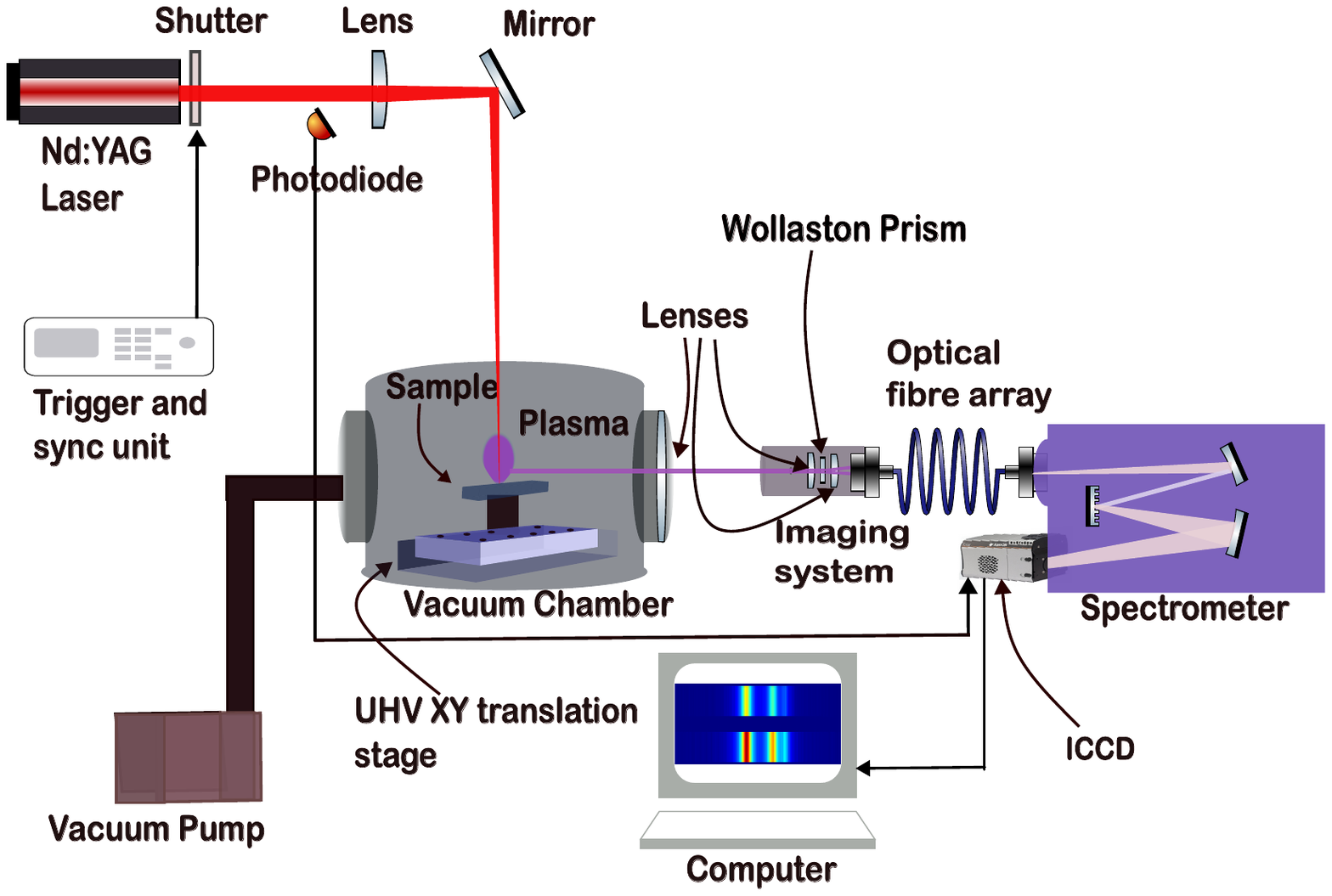}
		\caption{\label{fig:expsetup} Diagrammatic illustration of the experimental apparatus utilized in the present work. Nd:YAG laser is used to ablate aluminium sample kept inside the vacuum chamber. Schematic diagram of experimental setup showing the arrangement of the sample inside the vacuum chamber and imaging system along with the spectroscopic detection systems\cite{geethika}.}
	\end{figure}
	Figure \ref{fig:expsetup} depicts the schematic of the experimental setup employed for this study, consistent with our prior communication \cite{geethika}. In summary, the sample is affixed to a high-vacuum-compatible motorized translation stage kept within a high vacuum chamber which is pumped down using a Turbo molecular pump (TMP). A precision leak valve is employed to establish the background pressure at the desired value. After pumping down to the base value, the chamber is filled with nitrogen and maintained at 100 mbar in this investigation. A pulsed Nd:YAG laser with $\sim$ 10 ns pulse width operates at its fundamental wavelength and set for energy $\sim$ 150 mJ, loosely focused on the sample using a 60 cm plano convex lens to achieve an approximate spot diameter of $\sim$ 1 mm (laser fluence of $\sim 40 $ J/cm2), on an aluminium target to generate plasma.
	\par
	The plasma plume is imaged using a in-house developed imaging system with spatial resolution of 1 mm. A Wollaston prism is utilized to separate 
	horizontal (H) and vertical (V) polarizations into an optical fiber array of 600-micron diameter. The two polarizations are separated by the Wollaston prism into two identical spots of diameter 1 mm, separated by a distance of $\sim$ 2.5 mm. The fibers corresponding to each spot are coupled to the 1 meter spectrograph and binned together to create two channels (one for H polarization and the other for V polarization) on the spectrograph, allowing concurrent recording of the spectra. The imaging system is calibrated for complete visible wavelength range using a calibration lamp to ensure the accuracy of polarization separation and resolution.
	\par
	To mitigate the impact of statistical variations in the recorded intensity ratio between H and V polarizations, each spectrum is averaged over 20 ablations. The observed statistical fluctuations in recorded intensity ratio  with a spectral calibration lamp in visible wavelength range is well below 5\%. The imaging lens system is affixed to a precision translational stage, enabling systematic recording of emission along the propagation axis of the plasma plume.

	\section{Results and Discussion}\label{sec:results}	
	
	\begin{figure*}[!tbp]
		\includegraphics[scale=0.41,trim = {1mm 0 2cm 0},clip]{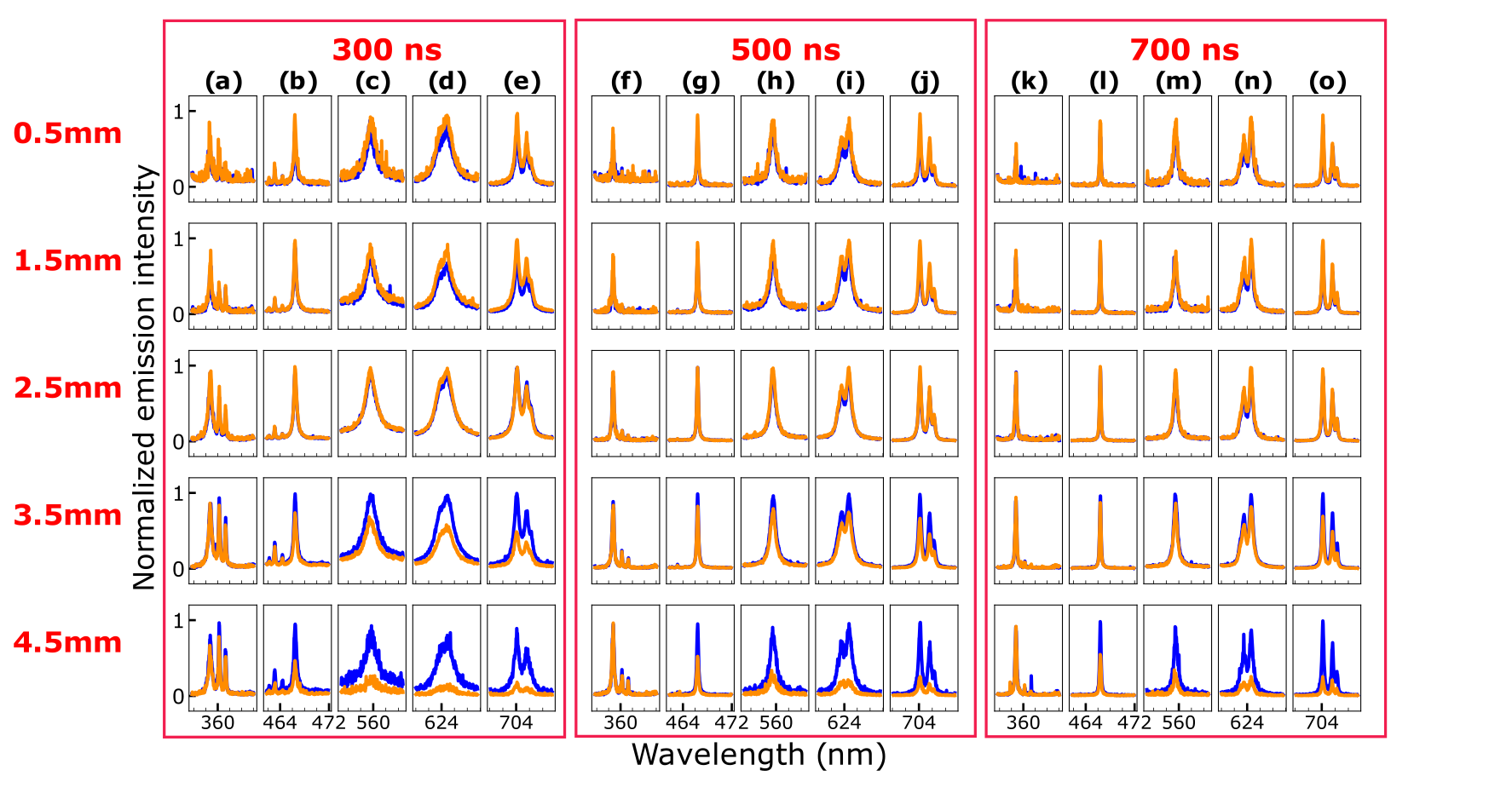}
		\caption{\label{fig:spectrum_var_pressure_3.5mm} Spectra of AL II emission lines peaking at different wavelength listed in table \ref{table:em_param} for horizontal (blue) and vertical (orange) polarizations. Column-wise distance from the sample increases. First five rows (a,b,c,d,e) shows the spectra at 300 ns after the ablation, next five rows (f,g,h,i,j) at 500 ns and last five rows (k,l,m,n,o) at 700 ns after the laser ablation. Each of the plot is normalized to the maximum intensity of emission observed among the two polarizations to optimize visibility.}
	\end{figure*}

	\begin{table}[hb]
		\caption{The spectral parameters describe the emission lines observed from various charge states of aluminum. Here, $g_k$ denotes the statistical weight, $A_{ki}$ represents the transition probability, and $E_k$ signifies the energy of the upper state involved in the respective transition. (taken from NIST database\cite{NIST_ASD})}
		\centering
		\begin{tabular}{ | p{1cm} | p{1.6cm} | p{1.5cm} | p{2.5cm} |p{1.5cm} |}
			\hline
			Ionic state & Wavelength (nm)  & g$_k$A$_{ki}$$\times$10$^8$ (s$^{-1}$) & Spectral Terms of Transition & E$_k$ (eV)\\ \hline
			Al II & 358.7 & 9.85  & $^3$F$_{0}$ $\rightarrow$ $^3$D$_{0}$ & 15.30 \\ \hline
			Al II & 466.30 & 1.74  & $^1$P$_{0}$ $\rightarrow$ $^1$D$_{0}$ & 13.26 \\ \hline
			Al II & 559.33 & 4.63  & $^1$D$_{2}$ $\rightarrow$ $^1$P$_{1}$ & 15.47 \\ \hline
			Al II & 623.17 & 4.20  & $^3$D$_{2}$ $\rightarrow$ $^3$P$_{1}$ & 15.06 \\ \hline
			Al II & 624.34 & 7.77  & $^3$D$_{3}$ $\rightarrow$ $^3$P$_{2}$  & 15.06 \\ \hline
			Al II & 704.21 & 2.89  & $^3$P$_{2}$ $\rightarrow$ $^3$S$_{1}$ & 13.08 \\ \hline
			Al II & 705.66 & 1.72  & $^3$P$_{1}$ $\rightarrow$ $^3$S$_{1}$ & 13.07 \\ \hline
			Al II & 706.36 & 0.573  & $^3$P$_{0}$ $\rightarrow$ $^3$S$_{1}$ & 13.07 \\ \hline
			Al III & 452.92 & 14.9  & $^2$D$_{5/2}$ $\rightarrow$ $^2$P$_{3/2}$ & 20.55 \\ \hline
			Al III & 569.66 & 3.51  & $^2$P$_{3/2}$ $\rightarrow$ $^2$S$_{1/2}$ & 17.82 \\ \hline
		\end{tabular}
		\label{table:em_param}
	\end{table}

	Initially we try to qualitatively study DOP of various line emissions from Al II, ranging from $\sim$ 350 nm to 700 nm.
	Figure~\ref{fig:spectrum_var_pressure_3.5mm} illustrates the evolution of intensities for both horizontal (H - blue) and vertical (V - orange) polarizations at different times and locations for various emission lines of Al II listed in table \ref{table:em_param}. From the figure it is evident that for emissions in the blue region (centred $\sim$ 360 nm) the polarization resolved intensities rather remains the same. 
	In the case of the emission centred at 704 nm, as reported earlier, at locations closer to the sample intensity for V polarization dominates for all the times and far away from the sample, intensity for H polarization starts dominating.
	
	\par
	
	\begin{figure*}[hbt]
		\includegraphics[height= 7 cm, width= 17 cm,trim = {2cm 0 4cm 0},clip]{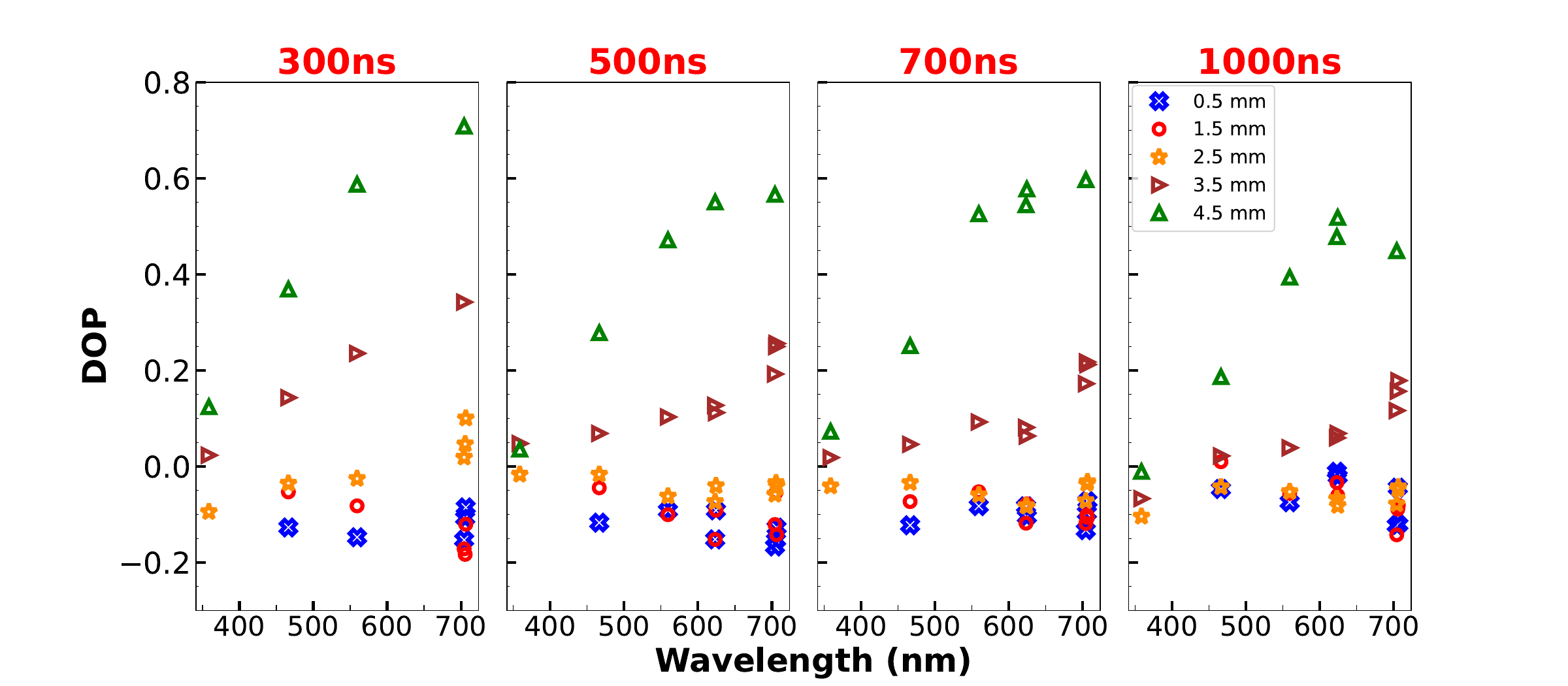}
		\caption{\label{fig:dop_al2} Variation of DOP with emission wavelength at different times after the laser ablation. Different points represents various locations within the plasma. The maximum error in the estimated DOP is expected to be around 10 \%.}
	\end{figure*}

	The extent of polarization is conventionally quantified utilizing the DOP\cite{SHARMA20073113,geethika}, as defined in equation \ref{eq:dop}.
	\begin{eqnarray}
		DOP = \frac{I_H-I_V}{I_H+I_V}
		\label{eq:dop}
	\end{eqnarray}
	where $I_H$ and $I_V$ are the intensities of H and V polarizations respectively. DOP of Al II lines listed in table \ref{table:em_param} are estimated for all spatial locations and delay times. Figure \ref{fig:dop_al2} depicts the variation in DOP with emission wavelength at different locations and times.
	Figure shows that at locations away from the sample (3.5 mm and 4.5 mm), DOP indeed increases with increase in emission wavelengths whereas, closer to the sample no such significant variation is seen. 
	The exact reasons for this wavelength-dependence of DOP is not fully understood, necessitating further experiments and theoretical investigations. Present study aims to highlight the implications of the observed wavelength dependence on estimation of plasma parameters like electron temperature and density which are crucial in the context of LIBS and PRLIBS.
	


	\begin{figure*}[t]
		\includegraphics[scale=0.43,trim = {4cm 0 4.2cm 0},clip]{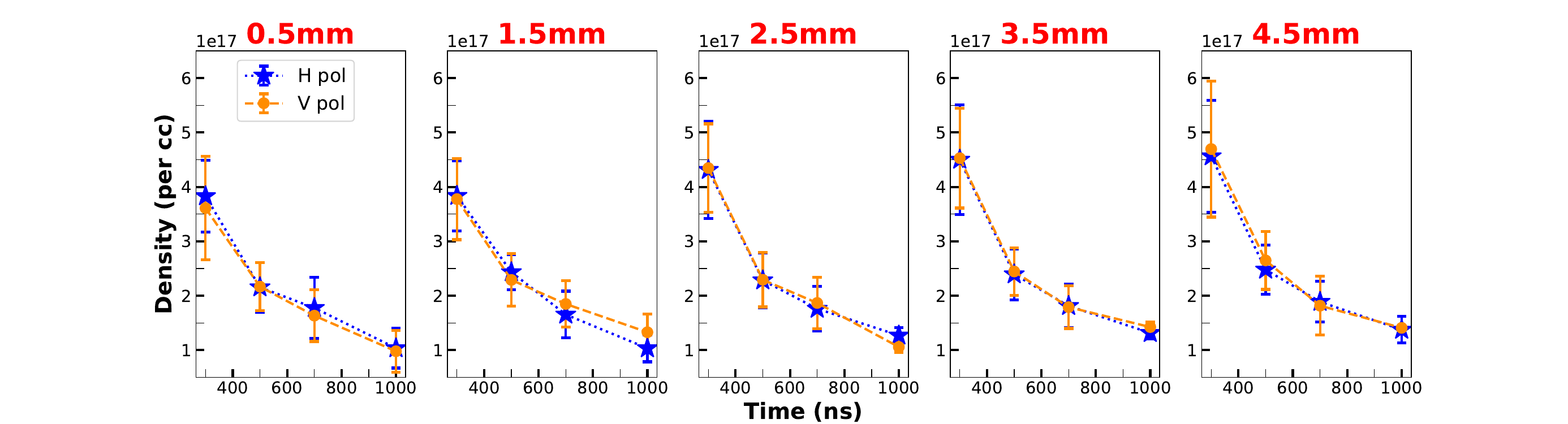}
		\caption{\label{fig:dens} Density variation of Al plasma with respect to time for different distances from the sample. The blue line shows H polarization and the orange line shows V polarization. Error bar is the statistical deviation in density estimated from emissions at 358.7 nm, 466.30 nm, 452.92nm, 569.66 nm, 559.33 nm and 704.21 nm .} 
	\end{figure*}

	Stark widths of emission lines are widely used for the estimation of electron densities in LPP\cite{jinto_pop,densharilal}, owing to relatively low electron temperature and high electron density. Other broadening processes like natural and Doppler broadening can be safely be avoided. However instrumental broadening has been considered in the calculations.
	The Stark broadening width, $\Delta \lambda_{1/2} (A^0)$, of emission lines can be expressed by equation~\eqref{Stark_eq}\cite{grieim}
	
	\begin{align}
		\Delta \lambda_{1/2} &= 2\omega\left(\frac{N_e}{10^{16}}\right)+  \label{Stark_eq} \\
		& \left[3.5A\left(  \frac{N_e}{10^{16}} \right)^{1/4}
		(1-1.2N_d^{-1/3}) \omega\left( \frac{N_e}{10^{16}}\right)\right] \notag
	\end{align}

	where $N_e (cm^{-3})$ and $ N_d$ are the electron density and number of particles in the Debye sphere ($N_d=1.72 \times10^9 \frac{T^{3/2}}{N_e^{1/2}}$) respectively,  $\omega$ and A are the electron and ion broadening parameters, which weakly depend on the plasma temperature. The second term is to account for ion collisions which can be neglected for heavier ions (non-hydrogenic). Hence the Stark broadening equation (equation \ref{Stark_eq}) can be shortened to 
	\begin{equation}
		\Delta \lambda_{1/2}=2\omega\left(\frac{N_e}{10^{16}}\right) A^0
		\label{Stark_eq_small}
	\end{equation}

	The plasma electron density is calculated using the reported Stark width parameters\cite{density559,density704,densal3} of  Al II and Al III emission lines  by estimating the Lorentzian width of both polarization independently. 
	Figure \ref{fig:dens} shows the variation of electron density with time for different locations inside the plasma. Average value for density was calculated using line emissions at 358.7 nm, 466.30 nm, 452.92nm, 569.66 nm, 559.33 nm and 704.21 nm. The standard deviation is plotted as the error bar. The obtained electron density was compared with the electron density calculated from the H-alpha line (due to trace of water) at locations and times where the signal was significant, and it was observed to fall within the error bars. The electron densities calculated from Al I emissions show significantly larger value than the values calculated from Al II, Al III and H-alpha lines indicating possibility of self- absorption for Al I lines.  Here it is important to note that the electron density estimated from both H and V polarizations is almost the same for all locations and times suggesting that the DOP has no effect on this estimation. Additionally, electron density calculated from emissions at 358.7 and 466.3 nm, the transitions showing smallest DOP, also falls within the error bar.
	These observations also point to the fact that Stark broadening of the emission line is not affected by polarization unlike the intensity. The figure shows that the electron density is higher than $1 \times 10^{17} cm^{-3}$ for all locations and times. The estimated electron density decreases over time in an exponential manner, but remains rather constant along the plume propagation direction. This trend can be attributed to the high background pressure\cite{Coons2011,diwakar,MAL2021112839}, as we discussed in our earlier work\cite{geethika}.
			
		
	\par

	In LPP, temperature determination typically employs techniques such as the Boltzmann plot method and line intensity ratio of same or subsequent ionic states, assuming local thermodynamic equilibrium (LTE) within the plasma as discussed in the introduction section. While using line intensity ratio of same charge states or Boltzmann plot, it is important to choose emission lines having large separation between upper energy levels to enhance the accuracy of temperature estimation. However, error in temperature estimated using line intensity ratio of two consecutive charge states is comparatively lower\cite{grieim}, as it takes into account the ionization potential in the estimation process. However, in this method, precise electron density information is essential. Boltzmann plot method, despite of its limitations, is extensively employed in LPP\cite{AGUILERA_bplot,BOUSQUET} because it simultaneously utilizes multiple lines from the same charge state to estimate plasma temperature, thereby eliminating the necessity for electron density information.
	\par
	In the case of higher electron density, as observed in LPP, where collisional processes dominates over radiative processes, the plasma is assumed to meet the condition of LTE\cite{densharilal}.
	Conventionally, LTE is verified with the help of McWhirter criteria\cite{ARAGON2008893} as given in equation \ref{Mc_whiter_condition}.
	\begin{eqnarray}
		N_e (cm^{-3})  \geq 1.6\times 10^{12} T^{1/2}(\Delta E)^3
		\label{Mc_whiter_condition}
	\end{eqnarray}
	where $T (K)$ represents the equilibrium temperature, $N_e$ represents the electron density,  and $ \Delta E (eV)$ denotes the energy difference between levels associated with the specific transition. From figure \ref{fig:dens} we can see that the electron density values are always greater than $1 \times 10^{17} cm^{-3}$ and thus can be considered to be in LTE for temperatures up to few hundreds of eV. 
	In such a scenario plasma temperature can be estimated using Boltzmann plot method\cite{ARAGON2008893} which uses the following expression (equation \ref{Boltzman_Plot})  
	\begin{equation}
		\ln{\bigg[\frac{I_{ij}\lambda_{ij}}{g_iA_{ij}}\bigg]}=\frac{-E_i}{k_BT_e} +C
		\label{Boltzman_Plot}
	\end{equation}   
	where $I_{ij}$, $\lambda_{ij}$,$ A_{ij}$, $g_i $, $E_i$ and $k_B$ are the spectral intensity, wavelength, transition probability, statistical weight of the upper state, upper state energy and Boltzmann constant respectively. 
	\par
	
\begin{figure}
		\begin{minipage}{0.5\textwidth}
			\centering
			\includegraphics[width=1\linewidth,trim = {0cm 0cm 2cm 0cm},clip]{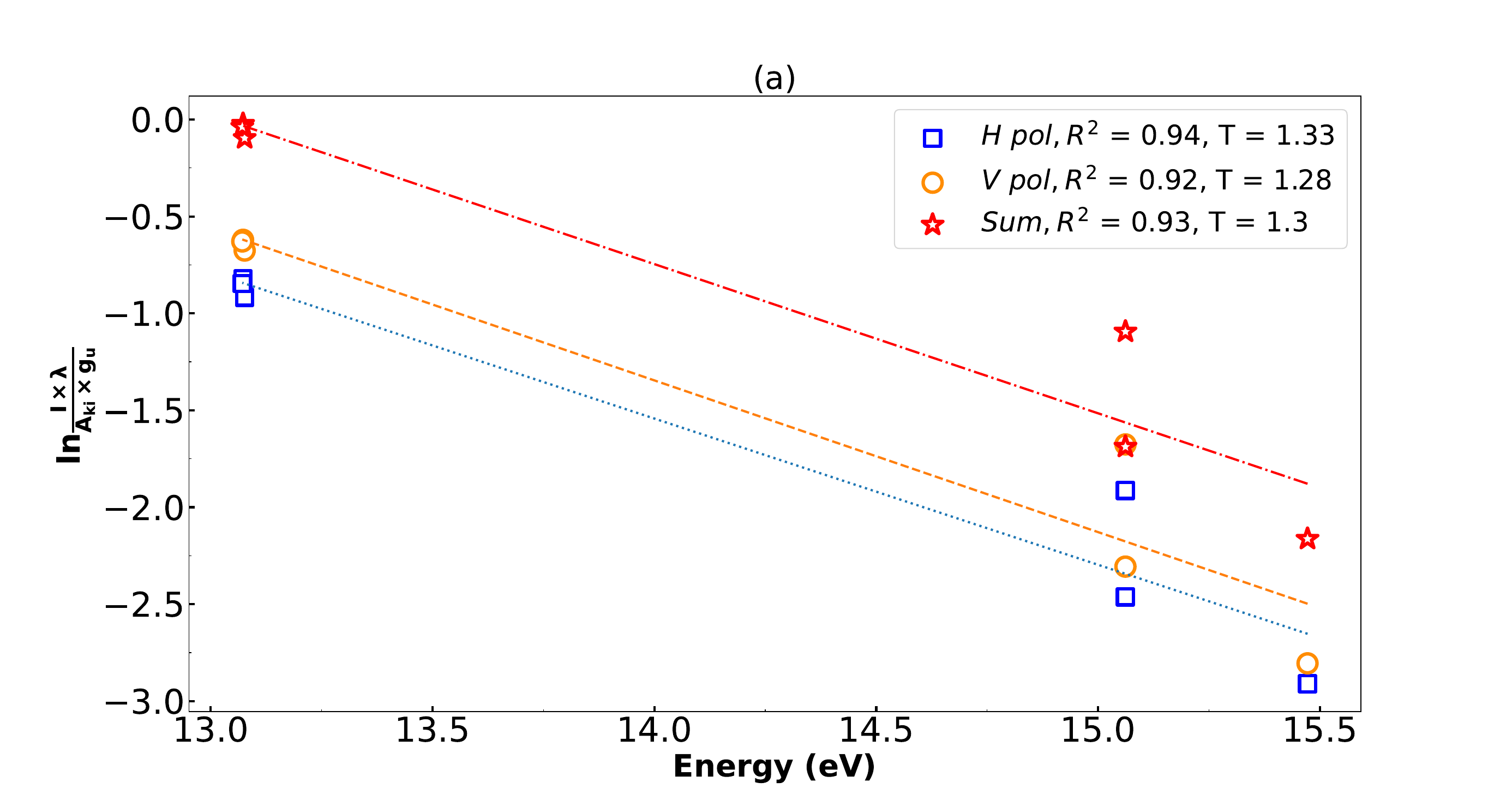}
			
			\label{fig:sub1}
		\end{minipage}
		\begin{minipage}{0.5\textwidth}
			\centering
			\includegraphics[width=1\linewidth,trim = {0cm 0cm 2cm 0cm},clip]{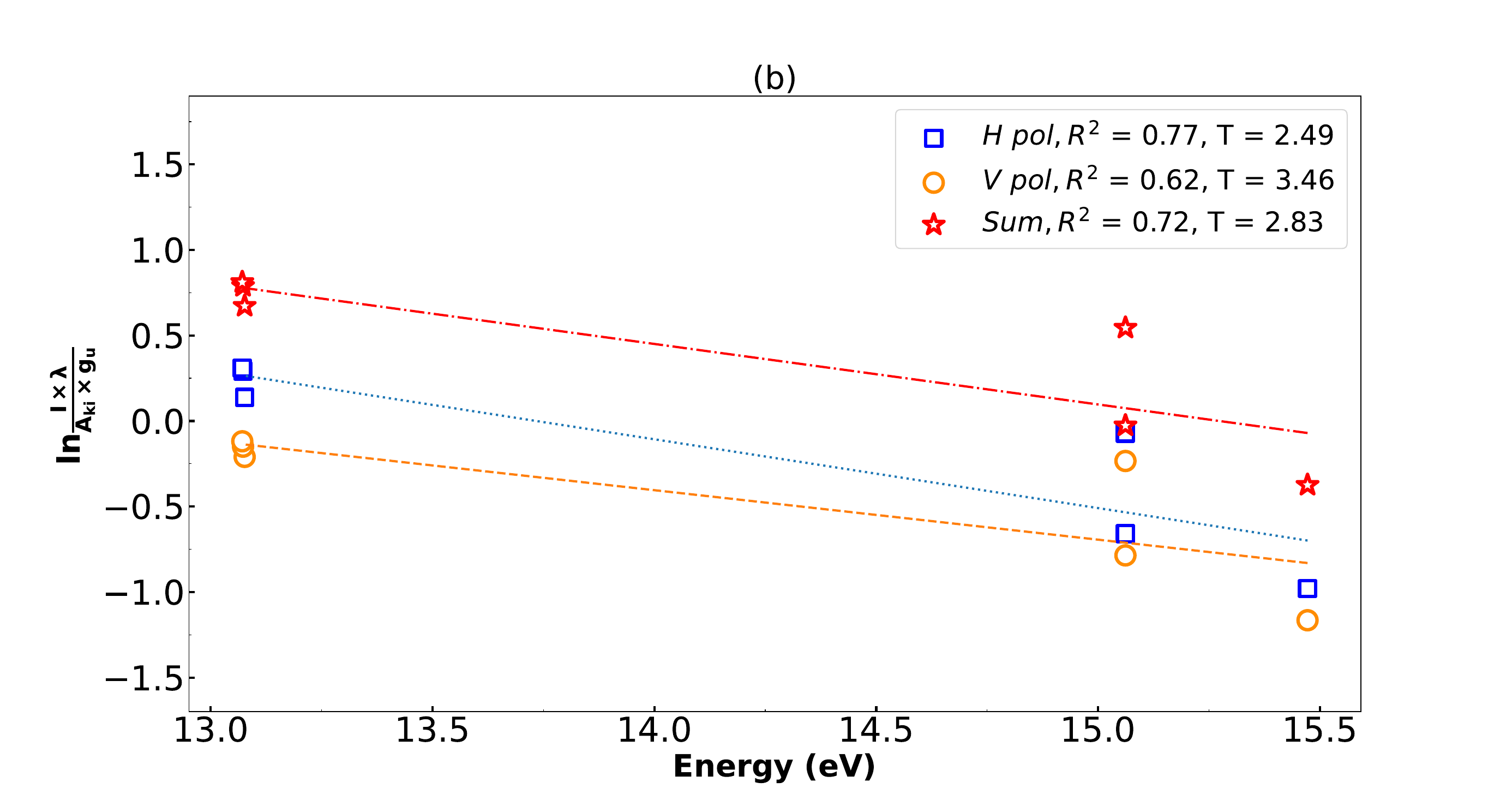}
			
			\label{fig:sub2}
		\end{minipage}
		
		\caption{Boltzmann plot at (a) 1.5 mm (b) 3.5 mm away from the sample and 700 ns after laser ablation. $R^2$ value of Boltzmann plot and estimated temperature for different polarizations is mentioned in the legend.}
		\label{fig:bplot_2pol}
	\end{figure}

	
	Temperature can be calculated by plotting the LHS of equation \ref{Boltzman_Plot} as a function of upper state energies and calculating the inverse of its slope. Al II lines 559.33, 623.17, 624.34, 704.22, 705.66 and 706.63 nm were used for the calculations. Multi-peak fit was employed to resolve and quantify the intensity of closely spaced line emissions. Additionally, the spectral response of the spectrograph is measured using standard light source to estimate the relative intensity variations of line emission across the wavelength range for the estimation of electron temperature.
	Temperature estimation was carried out separately for each polarizations. Figure \ref{fig:bplot_2pol} shows the Boltzmann plot for estimating temperature at a delay of 700 ns for both polarizations, as well as their sum for two locations: one with minimal DOP (1.5 mm) and the other with higher DOP (3.5 mm). At 1.5 mm the data fit well resulting in good $R^2$ value and reasonable temperature (with minimal deviation) for both the polarizations and their sum. Conversely, at 3.5 mm, the data points vary largely from the linear fit resulting in poor $R^2$ value. Also, the estimated temperature is substantially different for both the polarization and their sum. The sum of intensities of both polarizations yields a temperature between the two values estimated from the individual polarizations. The figure shows that the DOP plays a role in the accuracy of temperature estimation using Boltzmann plot method.
	\par
\begin{figure*}
		\includegraphics[scale=0.42,trim = {3.9cm 0 4.2cm 0},clip]{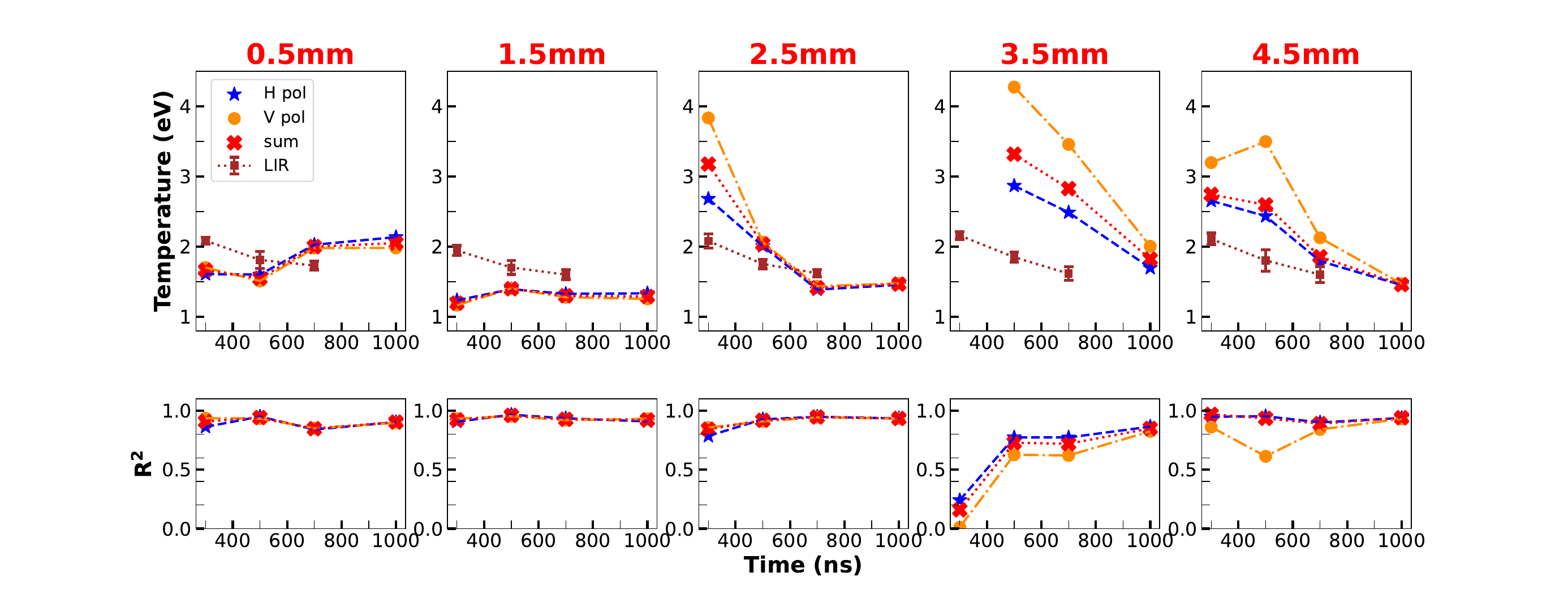}
		\caption{\label{fig:temperature_al2_bplot_lir} Time evolution of the temperature (first row) and $R^2$ value of Boltzmann plot (second row) of Al plasma at different distances from the sample (column-wise). Blue points show H polarization, orange points show V polarization and red points show the sum of the individual polarizations. Brown points show a the value of temperature calculated from the line intensity ratio (LIR) of successive charge states (averaged over all combinations of ratios between Al II and Al III lines mentioned in table \ref{table:em_param}).}
	\end{figure*}

 
	The temperatures at different locations and delay times are estimated for both the polarization using Boltzmann plot method. Figure \ref{fig:temperature_al2_bplot_lir} illustrates the variations in estimated temperature and the goodness of fit in the Boltzmann plot with time at different locations. Close to the sample (0.5 mm and 1.5 mm), the temperature estimate from both polarizations and its sum are very close for all delay times. The goodness of fit at these instances gives acceptable values (more than 0.9). However, at larger distances (3.5 mm and 4.5 mm) a noticeable difference in temperature is observed between the H and V polarizations. The goodness of fit also appears very poor.
	At 3.5 mm and 300 ns, the fit is extremely poor resulting in a very high value of the temperature and therefore it is excluded from the figure. A comparison of figure \ref{fig:dop_al2} and \ref{fig:temperature_al2_bplot_lir} shows that the temperature is found to differ noticeably for locations and times where the emission spectra exhibit higher DOP. However, for the location at which the polarization switching is observed, shows minimal difference between the temperatures estimated from individual polarizations.

	\begin{figure*}
		\includegraphics[scale=0.42,trim = {3.3cm 0 4.2cm 0},clip]{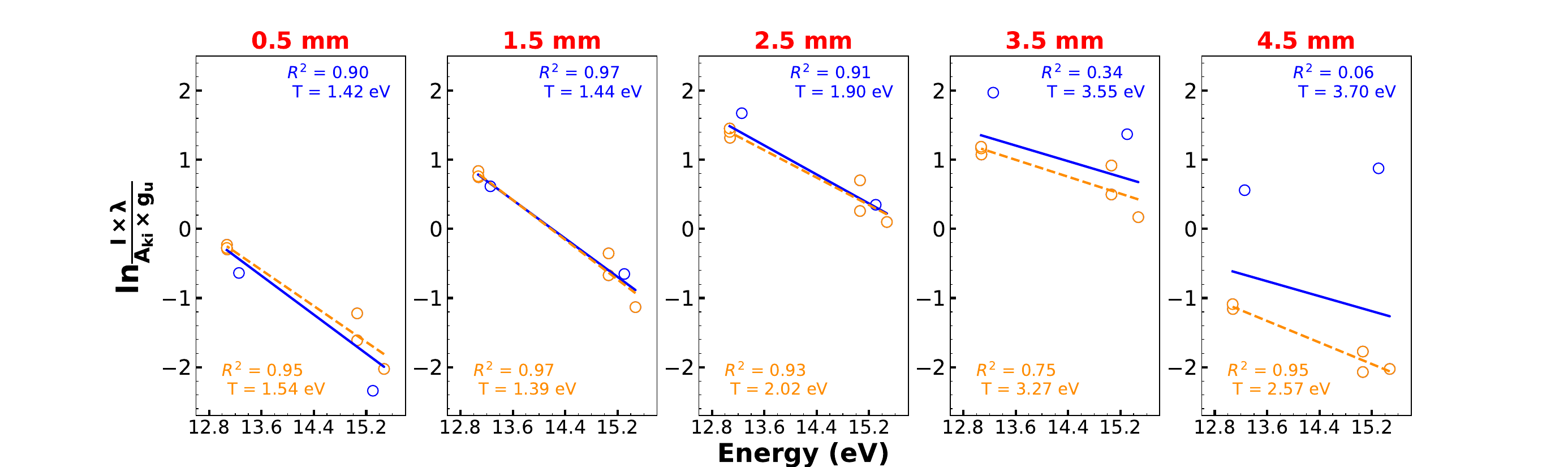}
		\caption{\label{fig:bplot_allpos_diff} Boltzmann plot for different positions inside the plasma plume at 500 ns after the laser ablation. Blue curve includes all the Al II lines included in table \ref{table:em_param} and orange line excludes lines at 466.3 and 358.7 nm. $R^2$ value of the fit and temperature from each plot are mentioned in each plot with (in blue, above graph) and without (in orange, below the graph) including the mentioned lines}
	\end{figure*}

	The accuracy of estimated temperature using the Boltzmann plot method depends on the separation of upper state energies of involved transitions. In present case, among the Al II transitions used for the estimation, the maximum separation between upper state energy is $\sim 2.4 eV$. At instances where high DOP is large, the estimated temperature is much higher than the maximum upper state energy separation of the transitions used. To confirm the accuracy of estimated temperature, we have also used line intensity ratio of successive ionic states\cite{grieim,densharilal}, which is considered as more accurate\cite{grieim}.
	The relation between the ratio of intensities of transitions from the successive ionized states and equilibrium temperature is given by equat ion \ref{Temp_int_R}\cite{densharilal}
	
	\begin{align}
		\frac{I'}{I} &=\frac{f'g'\lambda^3}{fg\lambda'^3}(4\pi^{3/2}a_0^3N_e)^{-1}\bigg(\frac{k_BT_e}{E_H}\bigg)^{3/2} \label{Temp_int_R} \\
		&\times exp\left(\frac{-(E'+E_{\infty}-E-\Delta E_{\infty})}{k_BT_e}\right) \notag		
	\end{align}
	
	here, $\lambda, E, g, f $ and $ I $ are the wavelength, upper level energy, statistical weight of upper level, oscillator strength and emission intensity respectively for the lower charge state. Likewise, $ \lambda', E', g', f' $ and $ I' $ are the mentioned quantities for higher charge state. $E_{\infty}$ denotes the ionization energy of the lower ionic state and $\Delta E_{\infty}$ represents its correction term. $E_H$ denotes the ionization energy of hydrogen atom, $a_0$, the Bohr radius and $k_B$ represents the Boltzmann constant. $N_e (m^{-3})$ stands for the electron density and $T_e$ (Kelvin) for the equilibrium temperature. Equation \ref{Temp_int_R} can be utilized to estimate the electron temperature at a particular instance, if the electron density values are known. Given the statistical variation in the estimated densities from  different spectral transitions is minimal (figure \ref{fig:dens}), it can be reliably used for the estimation of temperature at respective location and time.
	\par
	Figure \ref{fig:temperature_al2_bplot_lir} also includes the variation in temperature estimated using line intensity ratio method with time, for different positions in the plasma plume. Al II transitions at 559.33 and 704.21 and Al III at 452.92 and 569.66 nm were used to calculate the intensity ratios. Average value of the possible combinations are marked as temperature and the statistical deviation as error. At 1000 ns, Al III emissions were weak, thus not included in the plot. 
	The temperature estimated using line intensity ratio shows the expected trend with time but rather remains constant over the distance. This constancy in temperature with distance can be due to the confinement of plasma at high background pressure and plasma thermalization\cite{diwakar}.
	\par
	As the temperature estimated for both the polarizations using Boltzmann plot as well as  line intensity ratio shows a large deviation where the DOP is high, the validity of LTE needs to be verified using the Cristoforetti criteria \cite{CRISTOFORETTI201086} in addition to the McWhirter criteria. In case of inhomogeneous and transient plasmas (similar to our case) to be in LTE, Cristoforetti criteria\cite{CRISTOFORETTI201086,ishita2014} should also hold true. According to this condition during the time period of the order of relaxation time the diffusion length of atoms/ions must be less than the variation length of electron temperature and density. At position $x$ equation \ref{cristoforetti} should be satisfied	
	\begin{eqnarray}
		\frac{T(x)-T(x+ \lambda)}{T(x)}  \ll 1 \; ,
		\frac{N_e(x)-N_e(x+ \lambda)}{N_e(x)}  \ll 1
		\label{cristoforetti}
	\end{eqnarray}
	where T and $N_e$ are the electron temperature and density respectively. $\lambda$ is the diffusion length which is diffusion coefficient(D) times the relaxation time ($\tau_{rel}$). The expression for relaxation time ($\tau_{rel} \; (s)$) is given by equation \ref{rel_time} and equation \ref{diff_length} gives the diffusion length $\lambda \; (cm)$ \cite{CRISTOFORETTI201086}
	\begin{eqnarray}
		\tau_{rel} = \frac{6.3\times 10^4}{ N_e f_{ij}\langle \bar{g} \rangle}  \Delta E_{ij} (k_BT_e)^{1/2} exp \bigg(\frac{\Delta E_{ij}}{k_BT_e}\bigg)		
		\label{rel_time}
	\end{eqnarray}
	\begin{eqnarray}
		\lambda = 1.4 \times 10^{12} \frac{(k_BT_e)^{3/4}}{N_e}\bigg(\frac{\Delta E_{ij}}{ M_A f_{ij}\langle \bar{g} \rangle} \bigg) ^{1/2} exp \bigg(\frac{\Delta E_{ij}}{2k_BT_e}\bigg)		
		\label{diff_length}
	\end{eqnarray}
	where $N_e (cm^{-3})$, $k_BT_e (eV)$, $ \bar{g}$ and $M_A$ are the electron density, temperature, effective Gaunt factor and relative mass of the species under consideration respectively. $\Delta E_{ij} \; (eV)$ is the energy gap for a particular transition and $f_{ij}$ is its oscillator strength. For the minimum electron density  $\approx 1.1 \times 10^{17} (cm^{-3})$ and maximum temperature  $\approx 4.2 \; eV$, $\tau_{rel}$ is $ \approx 6 \times 10^{-11} s$ and $\lambda$ is $ \approx 5 \times 10^{-5} cm$ in case of Al II transition which has maximum energy gap ($ \Delta E = 2.21 eV, f_{ij} = 0.724 \cite{NIST_ASD}, \langle \bar{g} \rangle = 0.1$\cite{CRISTOFORETTI201086}). The variation in the electron density and temperature is negligible in the estimated range of relaxation time and diffusion length which means that the condition of LTE can be safely assumed. Additionally, given the background pressure of 100 mbar in our experimental configuration, the plasma can be confined for an extended duration, ensuring it remains in LTE. Hence, the observed deviation in estimated temperature using Boltzmann plot method is not arising due to the lack of LTE.
	\par
	One of the observed nature of DOP is its dependence on the emission wavelength(ref figure \ref{fig:dop_al2}). In the case of Boltzmann plot, different transitions across visible wavelength range are used for the temperature estimation. Therefore, the dependence of DOP on emission wavelength can effect the temperature estimated by this method.
	It is also important to mention that the temperature estimated after adding the intensities of both polarizations also deviates significantly  from the temperature estimated by the line intensity ratio method. This is indicative of a possible deviations of population of levels from the expected  Boltzmann distribution wherever the DOP is high.
	\par
	To understand the effect of wavelength dependence we estimated the plasma temperature using the sum of both polarizations of Al II transitions excluding those spectral lines in the blue region (358.7 and 466.3 nm). The derived temperature is then compared with the temperature obtained by incorporating these excluded transitions. Figure \ref{fig:bplot_allpos_diff} illustrates the Boltzmann plots with the inclusion (blue) and exclusion (orange) of the two blue wavelength transitions. The Boltzmann plot drawn after summing the H and V polarizations ensures the spectral intensity corresponds to the total emission. The temperature and goodness of fit are presented for both scenarios in their respective colours. From the figure, it is evident that the goodness of fit for both cases remains high for instances where the DOP is minimal. Moreover, the estimated temperature values are in good agreement, as expected. At 3.5 and 4.5 mm, where the maximum DOP was observed, a low $R^2$ value was obtained for the Boltzmann plot excluding the blue wavelengths. Further inclusion of the blue wavelengths worsens the fit as well as the estimated temperature.This clearly demonstrate that the emission intensities at blue wavelength region is not as per the expected Boltzmann distribution. Exact reason for this deviation is not clear, however it can be presumed that the cause of higher DOP and the deviation can be correlated.
	\par
	An additional potential cause for the considerable discrepancy in the estimated plasma electron temperature is self-absorption of the considered lines.  Given that the electron temperature here is within 2 eV and the density is approximately a few $10^{17} cm^{-3}$, a significant fraction of ions can be of Al II, which may undergo self-absorption. This phenomenon leads to erroneous estimates of line intensity in the absence of self absorption correction. In addition to affecting the estimated electron temperature, self-absorption can also have a significant impact on the estimated electron density. This is due to the erroneous estimation of the line width. The lines that are self absorbed will show a larger width than the actual Stark width\cite{ELSHERBINI}. However, as mentioned earlier the density estimated from different Al II, Al III and H-alpha lines is not significantly different, suggesting that self-absorption does not appear play a significant role in the present case.
	\par 
	The large deviation observed in the estimated temperature between the Boltzmann plot method and the line intensity method is of significant importance for PRLIBS and calibration free LIBS community. Normally, Boltzmann plot method is widely accepted for LIBS studies, however here it is clearly observed that when the emission is highly polarized (higher DOP), estimation of temperature using the Boltzmann plot method can be erroneous. This method estimates temperature around 4 eV for a plasma produced by a few GW ns laser, which is unlikely to be there\cite{Asamoah_temp,densharilal}. Similarly, close to the sample temperature does not follow the expected trend of fall with time, whereas line intensity ratio method shows a reasonable trend. The inaccuracies in Boltzmann plot method may be resulting from the processes involved in the observed anisotropy.
	Wubetu et.al\cite{wubetu,Wubetu_2020} highlighted this difference in temperature as a possible reason for anisotropy. However, we believe that the variation in the estimated temperature for different polarizations using Boltzmann plot method is likely due to the differences in the DOP of lines used to estimate the temperature. As mentioned earlier, DOP shows dependence on wavelength which results in some line emissions showing large variation in the intensity where the others not. This will likely result in some line emission intensities deviate from Boltzmann distribution\cite{ZHANG}, thus affecting the temperature estimation as shown in figure \ref{fig:bplot_allpos_diff} for different polarizations. Aghababaei et al \cite{aghababaei_time_evo} mentioned the importance of time resolved PRLIBS over the conventional PRLIBS. From our present work it can be inferred that, effective application of PRLIBS requires information of complete spatio-temporal evolution of DOP for the lines under consideration. Further, present study highlights the importance of the adopted method used for the estimation of electron temperature when there is significant variation in DOP.
	

\section{conclusion}
\label{sec:conclusion}

In the present study we demonstrate the effect of polarization on electron density and temperature for different polarizations estimated from optical emission spectroscopy (OES). Electron density remains same irrespective of polarization of the emission, as the line width remains the same. On the other hand, estimated electron temperature from Boltzmann plot method shows large variation with polarization, particularly depending on the wavelegth of lines used for estimating it. Temperature estimated from the line intensity ratio of two successive charge states appears to show less variation with polarization. The likely reason for this phenomenon appears to be deviation from the Maxwell-Boltzmann distribution of the energy states involved in the transition. This deviation suggests that the population of these energy states does not follow the expected thermal equilibrium distribution. We believe that present observations will have important implications in exploiting PRLIBS in elemental analysis.


\section{References}


%

\end{document}